\documentclass[iop]{emulateapj}
\usepackage{epstopdf}

\shorttitle{ALMA Detected Overdensity of Sub-mm Sources}
\shortauthors{Silva et al. 2015}
\begin{document}
\title{ALMA Detected Overdensity of Sub-mm Sources Around {\it WISE}/NVSS-selected $z$\,$\sim$\,2 Dusty Quasars }
\author{Andrea Silva\altaffilmark{1,2}, Anna Sajina\altaffilmark{1},
 Carol Lonsdale\altaffilmark{3}, \& Mark Lacy\altaffilmark{3}}
\altaffiltext{1}{Department of Physics and Astronomy, Tufts University, Medford, MA 02155, USA} 
\altaffiltext{2}{National Astronomical Observatory of Japan 2-21-1 Osawa, Mitaka, Tokyo 181-8588, Japan}
\altaffiltext{3}{National Radio Astronomy Observatory, 520 Edgemont Road, Charlottesville, VA 22903, USA}

\begin{abstract}
We study  the environments of  49  {\it WISE}/NVSS-selected dusty, hyper-luminous, z$\sim$2 quasars using the Atacama Large Millimeter/Sub-millimeter Array (ALMA) 345\,GHz images. We find that 17 of the 49 {\it WISE}/NVSS  sources show additional sub-mm galaxies within the ALMA primary beam, probing scales within $\sim$\,150\,kpc. We find a total of 23 additional sub-mm sources,  four of which in the field of a single {\it WISE}/NVSS source. The measured 870\,$\mu$m source counts are $\sim$\,10\,$\times$ expectations for unbiased regions, suggesting such hyper-luminous dusty quasars are excellent at probing high-density peaks. 
\end{abstract}

\newpage

\section{Introduction} \label{sec_intro}

Clusters are the largest gravitationally bound objects in the Universe. Finding clusters, especially at higher redshifts, is therefore critical both to constrain models of structure formation \citep[e.g.][]{poggianti10}, and to study the role of environment in galaxy evolution \citep[e.g.][]{peng2012}. However, finding high-$z$ clusters is challenging. For example, optical-color based techniques \citep{gladders05} rely on the red sequence galaxies which dominate the core cluster populations but only up to $z\sim$1.5 \citep{lidman2008,mei2009}.  At $z\sim$2, the Universe is only $\sim$3.3 Gyr old, i.e. insufficient time for a galaxy with a velocity of few hundred km/s to have crossed cluster-scale structures (few Mpc). Hence any overdense structure detected  would likely be protoclusters still in the process of virialization, hindering X-ray and Sunyaev-Zeldovich detection techniques. Spectroscopic and photometric redshift surveys that are deep enough to reach cosmologically interesting distances do not yet sample sufficient volumes to reach the largest possible clusters in a systematic way \citep{geach2011,yuan2014}.  We can avoid these issues by using strongly biased populations such as QSOs \citep{priddey2008,stevens2010,falder2011}, although see \citet{fanidakis2013} for an alternative view,  and radio galaxies \citep{wylezalek2013,dannerbauer2014} to find high-$z$ overdensities of star-forming galaxies.  For example, an excess of sub-mm galaxies \citep[SMGs; see][for a review]{blain2002} is observed in the fields of high-redshift radio galaxies (HzRGs) \citep{ivison2000, stevens2003, dannerbauer2014}. Since SMGs are believed to be the progenitors of local elliptical galaxies \citep{dunlop2001, smail2004, ivison2013} this excess is consistent with the view that we are observing proto-clusters at the time of build-up of their elliptical galaxy populations, with the central radio galaxy likely to evolve into the brightest cluster galaxy (BCG) \citep{miley2006}.  The higher angular resolution of ALMA opens the door for the first time to look for SMG overdensities in the near vicinity ($\lesssim$100\,kpc) of potential proto-cluster markers.  
 This  higher resolution has also shed light into the bright-end of the SMG population. For instance,  \citet{karim2013} found a significant deficit of source counts above $\sim$8 mJy compared with single-dish surveys, and concluded that, even at $S_{850\mu m}$\,$\gtrsim$4\,mJy, SMGs often reveal multiple distinct sources in higher resolution images.  
 
 In this letter, we compute the sub-mm galaxy source counts in the near vicinity of 49 {\it WISE}/NVSS-selected $z\sim$2, dusty, hyper-luminous (L$_{\mbox{\tiny IR}} \gtrsim$ 10$^{13}$ L$_{\odot}$), moderately radio-loud quasars (Lonsdale et al. 2015, submittted; hereinafter L15) in order to   study their  environments  as well as the potential effect of those rare, highly obscured quasars on their surroundings. 
 
Throughout this paper, we adopt {\sl Planck} cosmology values \citep{planck2014} of H$_{0}$=67.3 km s$^{-1}$  Mpc$^{-1}$, $\Omega_{m}$=0.315, and $\Omega_{\Lambda}$=0.685.

\section{Sample selection }\label{sec_sample}

Full details on the sample selection are given in L15.   The parent sample of 165 sources was selected on the basis of {\it WISE} 22\,$\mu$m and NVSS 1.4\,GHz detection, extremely red {\it WISE} [3.4]-[4.6]  and [4.6]-[12]  colors, and $\log(f_{20cm}/f_{22\mu m})>0$ (i.e. systems where the AGN dominates the radio emission). Their WISE colors imply obscured AGN typically at $z\gtrsim\,$1 \citep[see also][]{jarrett2011,yan2013}. The 22\,$\mu$m-detection implies rare, hyper-luminous ($L_{\rm{IR}}>10^{13}$L$_{\odot}$) galaxies. The sample further focuses on higher-$z$ sources by excluding optically-bright and extended sources.

The ALMA observed sub-sample of 49 was selected before the rest of the {\it WISE}/NVSS sample and differs slightly from the full sample (L15). In particular, it is limited to $\log(f_{20cm}/f_{22\mu m})<1$ sources avoiding radio-loud systems; its [3.4]-[4.6] colors are marginally redder implying dustier AGNs; and it reaches 22$~\mu$m flux densities that are 0.2\,dex fainter than the full sample. We examined the effect of the latter, by looking at our results if the fainter sources in the ALMA sub-sample are excluded and found no significant difference. Therefore, our results translate to the full {\it WISE}/NVSS sample, modulo the minority therein that are radio-loud (13\%) and/or have slightly bluer [3.6]-[4.5] colors (9\%).   Optical spectroscopic redshifts are available for 43 of the 49 ALMA sources and  range from $\sim$\,0.47\,--\,2.85, with a mean redshift of $\langle z\rangle$\,=\,1.69.

\begin{figure}[!htbp]
\begin{center}
\includegraphics[angle=0,scale=0.3]{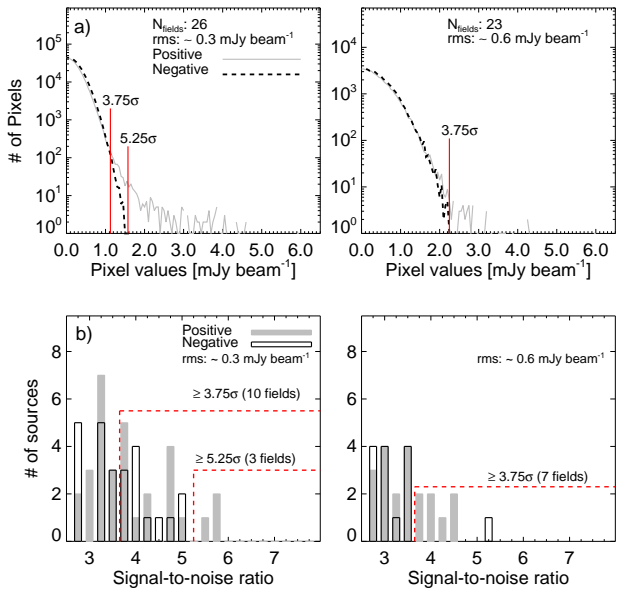}
\end{center}
\caption{{\it Top:} Pixel histograms for fields with different rms values before primary beam correction. 
 {\it Bottom:} Histogram of the signal to noise ratio of sources extracted in the ``positive" and ``negative" images in the 49 {\it WISE}/NVSS-fields.  
The threshold of SNR=3.75 adopted at which ``positive" sources exceed the number of ``negative'' sources is indicated with the red lines.  We also show the high-confidence cut (section \ref{sec_identify}).
 In parenthesis, we indicate the number of fields in which the sources with that threshold were detected.}
\label{fig_histogram_sources}
\end{figure}

\section{Observations \& Data Reduction} \label{sec_obs}

The ALMA observations  were  conducted in three epochs: twenty-three sources on November 16 2011, fourteen on May 25 2012, and twelve on August 28 2012. In each case, the observations were conducted in Band 7 (345\,GHz) with an 8\,GHz bandwidth. 
The time on source was $\sim$\,1.5\,min per object. The different number of antennae available (15, 19, and 23, respectively) led to different beam sizes and  rms values. Specifically, the synthesized beamsizes  are 1\farcs24, 0\farcs55, and 0\farcs45 for each run. By placing multiple apertures at random position in the images, we obtain the averaged rms  values for each run which are respectively: 0.60, 0.30, and 0.32 mJy beam$^{-1}$ and they do not significantly vary within the primary beam uncorrected for attenuation\footnote{The aimed rms was 0.5 mJy beam$^{-1}$, however as new antennae were added, the observations reached deeper rms. }.The primary beam size for ALMA at 345 GHz is 18\farcs2.  At the mean redshift of this sample,  $\langle z \rangle$=1.69, this primary beam size corresponds to $\sim$158 kpc.  The data were reduced using standard procedures and the Common Astronomy Software, {\sc casa} \citep{mcmullin2007}.  Twenty-six of the forty-nine {\it WISE}/NVSS sources were detected above a 3$\sigma$ level and none of them is resolved. 
For further details on the observations and data reduction see L15.

\begin{figure*}[!htbp]
\begin{center}
\includegraphics[angle=0,scale=0.80]{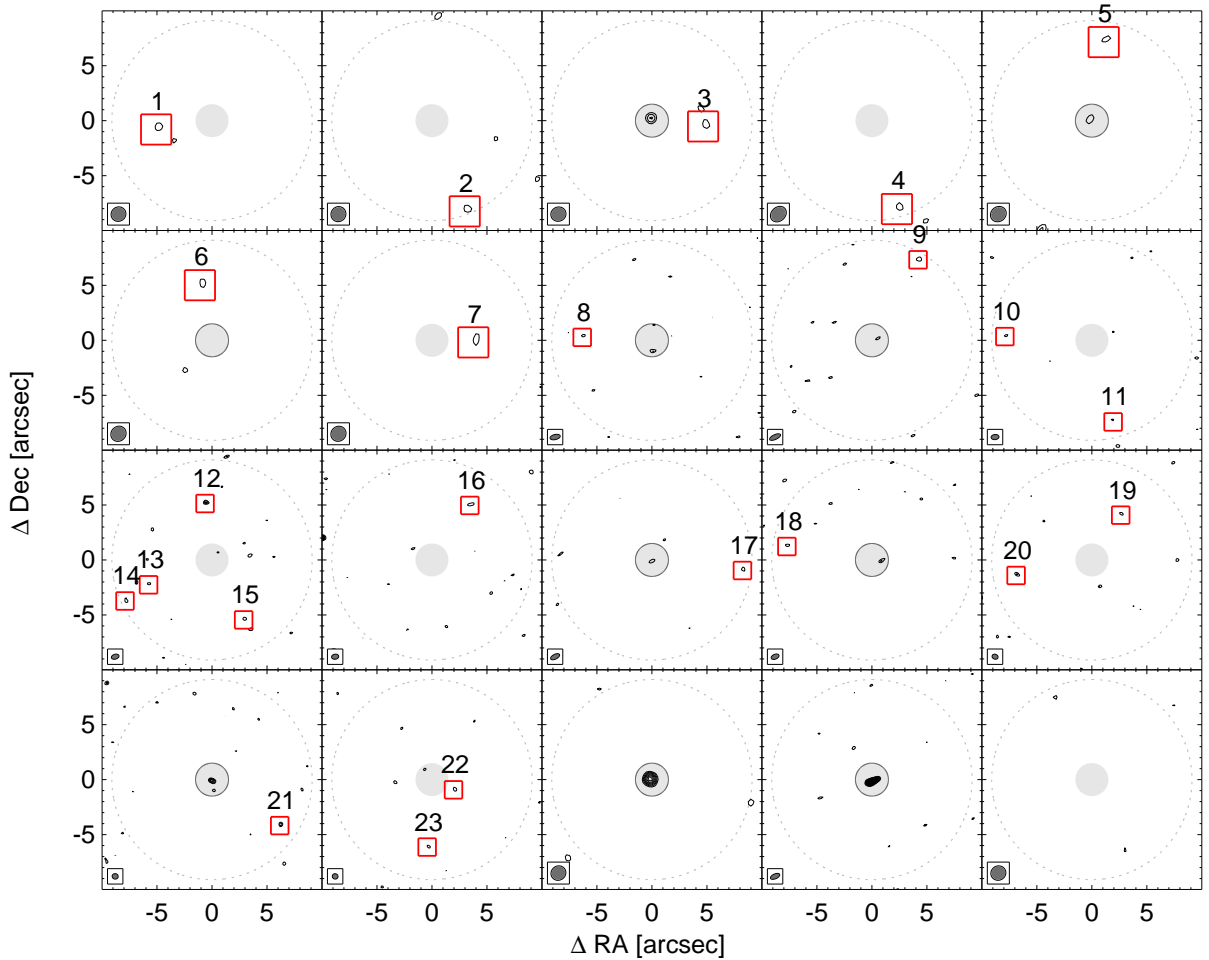}
\end{center}
\caption{870 $\mu$m continuum maps in which field sub-mm sources are  identified  around  the {\it WISE}/NVSS selected targets.  We include 3 fields with no additional source detection (bottom right in the image).  Contours start at 3$\sigma$ level and are in steps of 1$\sigma$. The identified sources with SNR$>$3.75 (Section \S  \ref{sec_identify}) are boxed and labeled with an Id number (Table \ref{tb_catalog}). The gray circles at the center of the images indicate the area where we do not search 
for sources, i.e. the location of the {\it WISE}/NVSS targets, and those with SNR$>$3 are indicated with the circle around the gray region. 
 The synthesized beam is shown at the bottom left of the images. The primary beam size  is indicated with the dashed circles.    \label{fig_sources}}
\end{figure*}
\section{Results}\label{sec_results}

\subsection{Identifying field sources}\label{sec_identify}

Figure\,\ref{fig_histogram_sources}{\it top} shows the pixel histograms, before primary beam correction, grouped by rms value. We include all  pixels  within the primary beam and outside a 1\farcs 5 radius from the center of the images to avoid the emission from the targeted {\it WISE}/NVSS sources.
The histograms show an excess of positive pixels starting at 3.75$\sigma$.   Although, a full P(D) analysis \citep[e.g.,][]{patanchon2009, glenn2010}  is beyond the scope of this paper, this excess confirms the presence of field sources. 

 We generate negatives of all the images and use  {\sc SExtractor} \citep{bertin1996} to find all $>$2$\sigma$ ``sources" in both our positive and negative maps. We only select sources within the primary beam.  Before primary beam correction, we measure  flux densities and associated uncertainties of all {\sc SExtractor}-selected sources using {\sc imfit} in {\sc casa}. Figure \ref{fig_histogram_sources}{\it bottom} shows histograms in signal-to-noise of these ``positive'' and ``negative'' sources. Beyond 3.75$\sigma$ the ``positive'' sources are in excess for both rms-groups. Using this threshold, we detect 23 sources in 17 fields\footnote{Eight of  these 17 {\it WISE}/NVSS sources present emission above 3$\sigma$ level, see Table \ref{tb_catalog}.}. We refer to this as our ``primary" serendipitous source sample. 
However, the spurious source fraction (as implied by the presence of ``negative" sources above this SNR threshold) is non-negligible. None of the images show obvious issues that may account for these such as insufficient cleaning.  We compute the spurious fraction as $N_{neg}/(N_{pos}+N_{neg})$ where $N_{neg}$ is the number of ``negative" sources above 3.75$\sigma$, and $N_{pos}$ is the number of ``positive" sources above this threshold.
 The result is a spurious fraction of 42\% and 12\% for images with rms of 0.3 and 0.6 mJy beam$^{-1}$, respectively.  Because of the high spurious fraction among the lower-rms fields, we also consider a more conservative cut where we only keep sources with SNR$>$\,5.25$\sigma$ in these lower-rms fields \footnote{A change in pixel size from 0\farcs1 to 0\farcs25 in the lower-rms fields (thus matching the higher-rms fields) does not affect our results.}.
This leaves us with a total of 10 sources spread among 10 fields. We refer to this as our ``high-confidence" serendipitous source sample.  In either case, this spurious source fraction is taken into account in computing the source counts (Section\,\ref{sec_scounts}).

Figure\,\ref{fig_sources} shows the fields with the 23 positive sources we identify in the primary serendipitous source sample (i.e. with SNR$>$\,3.75$\sigma$). The properties of the central {\sl WISE}/NVSS sources themselves are addressed in L15 (note 13 fields have no detections at all).
The detected sources are unresolved and none of them has been previously identified, based on a search on NED\footnote{Nasa Extragalactic Database http://ned.ipac.caltech.edu/}.  Their fluxes, uncorrected for primary beam attenuation, range from 1.56 to 3.11\,mJy\footnote{We apply a primary beam correction using
$F_{pbc}=F_{0}\exp\left(4\ln(2)\frac{d^2}{\theta^2}\right)$, where $F_{pbc}$ is the corrected flux, $F_{0}$ is the uncorrected flux, $\theta$ is the primary beam size, and $d$ is the distance of the source from the center of the image.}.  
 The key parameters for all the sources are presented in Table \ref{tb_catalog}. 
 Based on blank-field counts, we expect $\sim$\,0.04(0.02) sources/field in the rms$\sim$\,0.3(0.6)\,mJy fields, respectively. Without any further analysis, our observations show that 30\% of fields have at least 1 serendipitous source already implies counts that are $\sim$\,10\,$\times$ in excess of blank-sky sub-mm counts. The existence of fields without serendipitous detections, is consistent with our estimated counts which imply a probability of finding a source to be $<$\,1 in any given field.

\subsection{Angular distribution of serendipitous sources}\label{sec_angdist}

\begin{figure}[!htbp]
\begin{center}
\includegraphics[angle=0,scale=0.50]{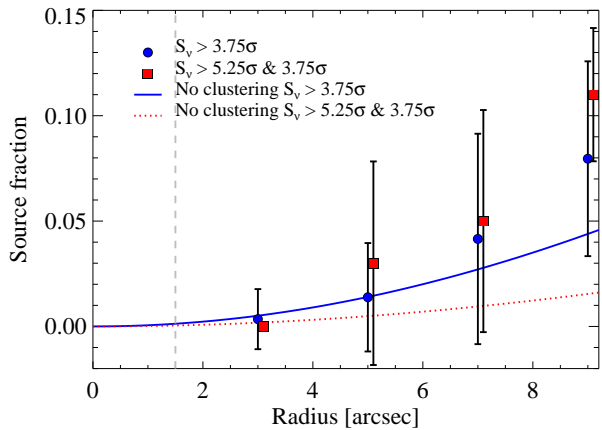}
\end{center}
\caption{ Circles and squares (shifted to the right by 0\farcs1) represents the cumulative fraction of detected sources in each field within different radii obtained for the different selection thresholds (as indicated).  The solid and dotted lines indicate the expected number of serendipitous sources if they are randomly located with no angular clustering.  The interior limit of source detection is 1\farcs5. }
 \label{fig_sfraction} 
\end{figure}

 For each field, we calculate the cumulative number of detected sources at different angular radii. We plot the mean of these cumulative fractions and compare with the expected fraction of sources with no angular clustering (Figure \ref{fig_sfraction}). 
 For the primary serendipitous sample, we find no evidence of angular clustering, which would manifest as an excess of sources toward the central source, relative to a random distribution. This is consistent with  the result of \citet{jones2015},  
  toward a sub-sample of our same parent population of 30 red {\it WISE}/NVSS sources observed with SCUBA at 850 $\mu$m. They 
 sample angular scales that start at our external radius and extend up to 1.5\,arcmin. Our high-confidence sample even shows a tentative sign of a dearth of SMGs in the vicinity of the {\it WISE}/NVSS sources. This may be the result of feedback effects from the central source quenching star-formation in the near vicinity. However, given our error bars, the significance of this result is only $\sim$\,2$\sigma$ and needs further investigation before it is conclusive.  
 
\begin{deluxetable*}{ccccccccccc}[!htbp]
\tabletypesize{\scriptsize}
\tablecaption{ Basic properties of the SMGs identified in the fields of the 49 {\it WISE}/NVSS selected targets. The fields are separated by high (top) and low rms values (bottom) \label{tb_catalog} }
\tablewidth{0pt}
\tablehead {{\it WISE/NVSS} & Id in  & R.A. (J2000)  & Dec. (J2000) & S$_{870\mu m}$\tablenotemark{b}  & S/N &D\tablenotemark{c} & $z\tablenotemark{d} $ &D\tablenotemark{e} & A$_{T}$& Detect.\tablenotemark{f}\\
  Field Name\tablenotemark{a}   & Fig. \ref{fig_sources} & (h:m:s)  & (d:m:s) & (mJy) &  &(\arcsec)& & (kpc)& (arcmin$^{2}$)&  }   \startdata
 (J035448.24$-$330827)& 1 &03:54:48.62&$-$33:08:27.70&2.29$\pm$0.58&3.98&   5.1 & 1.373&44& 0.25& No\\  
(J051905.84$-$081320)  &2 &05:19:05.62& $-$08:13:28.62 & 4.53$\pm$1.13 &4.31&  8.8&2.000 & 76&1.62& No \\
(J053622.59$-$270300) &3 & 05:36:22.24& $-$27:03:00.73 & 2.47$\pm$0.69& 3.76&4.7 & 1.791&41&0.37& Yes \\
(J061405.55$-$093658) &4&06:14:05.37&$-$09:37:06.57 &4.23$\pm$1.07 &4.05& 8.3& 2.000 &71&1.56& No \\
(J063027.81$-$212058) &5& 06:30:27.72&$-$21:20:51.39 &4.81$\pm$1.08 &4.60&  7.2& 1.439 &42& 1.65& Yes\\  
(J064228.93$-$272801) &6& 06:42:29.00&$-$27:27:56.61 &3.15$\pm$0.84 &3.80 & 5.1& 1.340 &44&0.97& Yes \\  
 (J070257.20$-$280842) &7& 07:02:56.89&$-$28:08:42.30  &   3.22$\pm$0.70 & 4.73 & 3.7& 0.943 &30&1.01& No \\ \\ \hline \\
 J130817.00$-$344754 & 8&13:08:17.52 &$-$34:47:53.14  &1.83$\pm$0.54(2.12) &3.86 &6.3 & 1.652& 55&1.25& Yes \\
(J143419.59$-$023543) &9&14:34:19.30 & $-$02:35:36.48 & 2.92$\pm$0.63 &5.77 &  8.4& 1.922 &72&1.86& Yes \\  
J143931.76$-$372523 & 10&14:39:32.42 &$-$37:25:23.25  &1.71$\pm$0.49(1.98) &3.85 &7.9 & 1.200&67&1.06& No \\
J143931.76$-$372523 & 11& 14:39:31.60&$-$37:25:30.91  &2.06$\pm$0.60(2.27) &4.80 &7.7 & 1.200& 66&1.55& No \\
J151003.71$-$220311 & 12&15:10:03.76 &$-$22:03:04.51  &1.30$\pm$0.38(1.59) &3.75 &5.2 & 0.950&42&0.37& No \\
J151003.71$-$220311 & 13&15:10:04.14 &$-$22:03:11.90  &1.87$\pm$0.45(2.11) &4.91 &6.2 & 0.950&50&1.29& No \\
J151003.71$-$220311 &14 & 15:10:04.28&$-$22:03:13.51  &2.33$\pm$0.65(2.70) &4.44 &8.7 & 0.950& 71&1.74 & No\\
J151003.71$-$220311 & 15& 15:10:03.50&$-$22:03:15.26  &1.68$\pm$0.39(1.92) &4.40 &6.2 & 0.950& 51&0.99 & No\\
J151424.12$-$341100 & 16&15:14:23.84 &$-$34:10:55.03  &1.78$\pm$0.49(1.99) &4.81 &6.0 & 1.080& 50&1.15& No\\
(J152116.59+001755) &17& 15:21:16.03& +00:17:54.20& 3.04$\pm$0.82 &5.86&  8.3& 0.700 &61&1.87 & Yes\\ 
J154141.64$-$114409  &18 &15:41:42.17 &$-$11:44:07.85  &1.82$\pm$0.61(2.22) &3.76 &7.8 & 1.580& 68& 1.22 & Yes\\
(J163426.87$-$172139) &19&16:34:26.67 &$-$17:21:36.05  & 1.90$\pm$0.44(2.03) &5.57 & 4.8&2.070 & 41 & 1.36 & No\\
J163426.87$-$172139 & 20&16:34:27.34 &$-$17:21:41.47  &2.15$\pm$0.47(2.36) &5.07 &7.0 & 2.070&60 & 1.62 & No\\
J164107.22$-$054827 & 21&16:41:06.80 &$-$05:48:30.97  &1.78$\pm$0.53(2.14) &3.91 &7.4 & 1.830& 64& 1.15 & Yes\\
J170204.65$-$081108 & 22& 17:02:04.51&$-$08:11:08.51  & 1.65$\pm$0.44(1.80) &4.80 &2.2& 2.850& 18& 0.97& No\\
J170204.65$-$081108 & 23& 17:02:04.65& $-$08:11:13.95  &1.88$\pm$0.60(2.09) &4.13 &6.1 & 2.850& 49&1.35 & No
\enddata
\tablenotetext{a}{In parenthesis the sources selected with the high-confidence level (Section \ref{sec_identify}). }
\tablenotetext{b}{De-boosted and primary beam corrected flux density. Errors are obtained by adding in quadrature the errors obtained in {\sc imfit} and the rms values at the position of the serendipitous source after primary beam correction. The boosted fluxes are shown in parenthesis.}
\tablenotetext{c}{Angular distance of the source from the target {\it WISE}/NVSS source's position.  }
\tablenotetext{d}{Redshift {\it WISE}/NVSS target.}
\tablenotetext{e}{Physical separation between the new detected SMG and the {\it WISE}/NVSS target  assuming they are at the same redshift.}
\tablenotetext{f}{Detection {\it WISE}/NVSS target above 3$\sigma$ level.}
\end{deluxetable*}

\subsection{Source Counts} \label{sec_scounts}

We calculate the integral source counts $N(>S)$ for the additional sub-mm sources by following the method described in \citet{ono2014}:
\begin{equation}
N(>S)=\sum_{S_{i}>S}   \frac{1-f_{c}(S_{i})}{C(S_{i}) A_{\mbox{\tiny T}}(S_{i})} .
\end{equation}

The fraction of spurious sources $f_{c}(S_{i})$  is 12\% and 42\%, for our primary serendipitous source sample,  which we treat as a constant for fields with the same rms values
The completeness $C(S_{i})$ is calculated by injecting 50 artificial sources at random positions in an  image where all sources with SNR$\geq$3.0 are removed. This is performed for two images representing the two different rms values. The procedure is repeated 1000 times and then completeness is computed from the fraction of recovered sources at different flux densities. 
Above 3$\sigma$ completeness ranges from 85-100\% and is applied before primary beam correction.
The effective area is the area in which a source with intrinsic flux density $S_{i}$ will be detected in one field. The total effective area $A_{\mbox{\tiny T}}(S_{i}) $ is the addition of the effective areas in all the fields with similar rms values, i.e. for a given flux density we add the effective areas of all fields with either rms of 0.3 or 0.6 mJy beam$^{-1}$. Their values are presented in Table \ref{tb_catalog}.
We estimate the contribution due to flux boosting by measuring the flux densities of the injected sources used to calculate completeness $f_{out}$  and take the ratio with their assigned fluxes $f_{in}$ and check their variation as function of SNR. Flux boosting is negligible above 3$\sigma$ for sources in high-rms fields. For sources in the lower-rms fields, it ranges from $f_{out}/f_{in}:$1.42-1.05 for sources with SNR from 0.5-6.5$\sigma$.

\begin{figure}[!htbp]
\includegraphics[angle=0,scale=0.35]{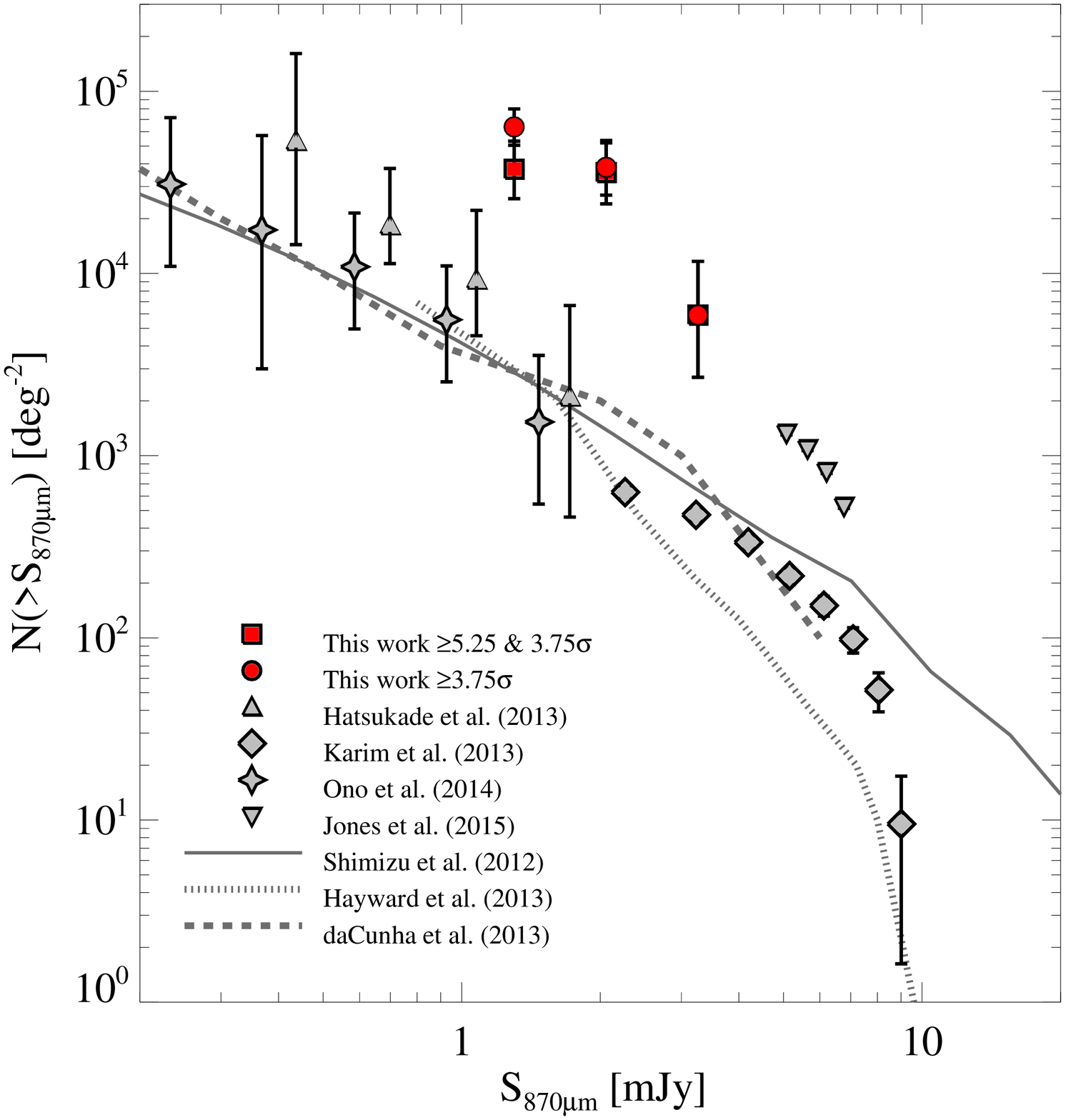}
\caption{ Integral source counts of SMGs around extremely red {\it WISE}/NVSS-sources (red circles) obtained with ALMA observations  at 870 $\mu m$ with SNR$>$3.75. We also plot the counts obtained using the high-confidence limit (squares). 
We overplot the ALMA counts  determined by \cite{hatsukade2013}  at 1.3mm,  \cite{karim2013} at 870$\mu$m, and \citet{ono2014} at 1.2 mm,  and the  models of \cite{shimizu2012}, \cite{dacunha2013}, and \cite{hayward2013}.   In addition, we include the counts obtained by \cite{jones2015} toward a subsample of the same parent population of our targets. All of these results were scaled to 870 $\mu$m using a modified blackbody.  
For the \citet{karim2013} counts we applied a correction factor of 2$\times$ underdensity of SMGs in the LESS field \citep{swinbank2014}.
Our counts are $\sim$\,25 to 10\,$\times$ stronger than the comparison models and observations toward blank fields and in agreement with the counts of \cite{jones2015}. } 
\label{fig_scounts}
\end{figure}

Figure \ref{fig_scounts} shows the source counts obtained from the primarily serendipitous source sample (i.e. with SNR$>$3.75) and from the high-confidence sample (i.e. with SNR$>$5.25 in the lower-rms  fields). The results are consistent with each other suggesting the details of the serendipitous sample selection do not significantly affect our conclusions. We compare our counts with those expected from models \citep{shimizu2012, hayward2013, dacunha2013},  previously measured ALMA counts obtained for presumably unbiased populations \citep{karim2013, hatsukade2013, ono2014}, and also with the results of \cite{jones2015}.
These literature counts are converted to counts at 870\,$\mu$m by using a modified blackbody \citep[as in][]{hatsukade2013,ono2014} assuming $\beta$\,=\,1.5, $T_{d}=35K$ and $z=2.5$, which are typical values for SMGs \citep{coppin2008,yun2012}. 
Our counts are significantly in excess of both models and observations for field SMGs. However, within the uncertainties, they agree with the counts obtained by \cite{jones2015}, who find an excess of 6\,$\times$ relative to blank-fields on scales of $\lesssim$ \,1\,Mpc. 
Our counts imply an even stronger excess of $\sim$\,10\,$\times$ relative to blank-sky surveys and are on much smaller spatial scales  compared with \cite{jones2015} ($<$\,150\,kpc).  Compared with the closest model \citep[that of][]{dacunha2013}) this excess is at the $\sim$\,5$\sigma$ level. 

\section{Discussion}\label{sec_discussion}

What does an overdensity of $\gtrsim$\,10\,$\times$ imply for our WISE/NVSS-selected  $z$\,$\sim$\,1.7 quasars? This is even stronger than the overdensity of 6\,$\times$ around a comparable sample found in \cite{jones2015} on scales of $\lesssim$\,1\,Mpc. Since the counts of \cite{jones2015} could be affected by unresolved sources \citep[][]{karim2013}, without flux density overlap, we cannot assess whether or not the small difference in level of overdensity as measured in this paper and in \cite{jones2015} is significant. The drop in overdensity from scales of $<$\,150\,kpc to nearly 1\,Mpc as sampled between the two papers is far weaker than expected based on local structures. For example, \citet{budzynski2012} show that the surface density of galaxies in local groups and clusters drops by $\sim$\,100$\times$ from roughly 100\,kpc to 1\,Mpc. This implied lack of significant clustering is also consistent with the angular distribution of sources as seen in both our paper and in \cite{jones2015}. 

We looked for trends in presence of serendipitous sources in a field vs. redshift, total luminosity,  $870~\mu$m flux (or just sub-mm detection of the central {\it WISE}/NVSS source), and radio power. We found no significant trends with respect to any of these properties of the central dusty quasar. However, as discussed in Section\,\ref{sec_sample}, the ALMA sub-sample is lacking the most radio-loud sources in the parent {\it WISE}/NVSS sample and therefore the range in radio power probed may be too small to detect any trends with the strength of the radio AGN.

Our galaxies are rare, hyper-luminous, obscured quasars, with significant dust masses\footnote{Assuming all the 870$\mu$m emission is thermal.} (see L15) and by extension cold gas masses. This is similar to findings of HzRGs including the Spiderweb \citep{ivison2012, emonts2013}. The large dust masses imply young objects that will likely evolve into red and dead ellipticals. This study, as well as \cite{jones2015}, suggest these sources reside in significantly overdense regions, but not yet fully-formed clusters. The observed overdensity is in SMGs, which implies significant star-formation, again similar to the much more radio-powerful Spiderweb galaxy \citep{dannerbauer2014}. We show tentative signs of a dearth of gas-rich star-forming galaxies at the very centers of these potential proto-clusters -- potentially an early indication of morphological segregation in clusters. Upcoming {\it Spitzer} IRAC imaging of these regions will help further explore this issue, as the mass-selected IRAC sources should show a more centrally-concentrated angular distribution.

This study demonstrates the utility of environmental studies using archival ALMA images, which allow us to sample at high resolution the dense cores of potential proto-clusters. Follow-up redshift studies are needed to confirm if we have indeed detected proto-clusters, especially in the case of J151003.71$-$220311 which shows the highest overdensity of SMG sources.

\section{Conclusions}\label{sec_conclusion}

We examined the near fields of 49 {\it WISE}/NVSS-selected dusty, hyper-luminous quasars at $\langle z\rangle$\,$>$\,1.7, using ALMA 870 $\mu$m images.  We found 23 additional SMG sources in 17 of these 49 fields. 
These imply a source density $\sim$10\,$\times$ higher than expectations for field SMGs, consistent with previous studies in the fields of $z$\,$\sim$\,2 QSOs and radio galaxies\citep{venemans2007, kodama2007, priddey2008, stevens2010, matsuda2011, husband2013,   dannerbauer2014}.  Our results are consistent with \cite{jones2015}, although we focus on smaller spatial scales, sampling the dense inner cores of potential proto-clusters. 

\acknowledgements

We are grateful to B. Hatsukade, and Y. Matsuda for useful discussions and A. Blain for a critical reading of an earlier draft of this manuscript.  We also thank 
Y. Ono, I. Shimizu, and E. da Cunha for providing their counts and models.  We thank the anonymous referee for useful feedback which helped improve this paper. A.S. is partially supported by NSF AST-1313206. This paper makes use of the following ALMA data: \dataset{ADS/JAO.ALMA\#2011.0.00397.S}. ALMA is a partnership of ESO (representing its member states), NSF (USA) and NINS (Japan), together with NRC (Canada) and NSC and ASIAA (Taiwan), in cooperation with the Republic of Chile. The joint ALMA Observatory is operated by ESO, AUI/NRAO and NAOJ. The National Radio Astronomy Observatory is a facility of the National Science Foundation operated under cooperative agreement by Associated Universities, Inc. Cosmology calculations in this paper were performed using the Ned Wright Cosmology Calculator.

\end{document}